\pdfoutput=1
\documentclass[10pt,conference]{IEEEtran}
\IEEEoverridecommandlockouts
\usepackage{amsmath,stfloats}
\usepackage{graphicx}
\include{psfig}
\usepackage{epsfig}
\usepackage{array}
\usepackage{psfrag}
\usepackage{amssymb}
\usepackage{mathdots}
\include{psfig}
\usepackage{color}
\usepackage{arydshln}
\usepackage{psfrag}
\usepackage{gensymb}

\usepackage{mathtools}
\DeclareMathOperator{\ExpOp}{E}
\DeclarePairedDelimiterX{\ExpArg}[1]{[}{]}{#1}
\newcommand{\Exp}{\ExpOp\ExpArg*}
\DeclareMathOperator{\Tr}{Tr}
\usepackage{amsthm}

\usepackage{amsmath}
\usepackage{amssymb}
 
\usepackage{breqn}
\usepackage{graphicx}
\usepackage[hidelinks]{hyperref}
\renewcommand\IEEEkeywordsname{Index Terms}
\usepackage[utf8]{inputenc}

\usepackage{epstopdf}

\makeatletter
\def\footnoterule{\kern-3\p@
  \hrule \@width \linewidth \kern 2.6\p@} 
\makeatother

\begin{document}
\date{}  
\title{Receive Spatial Modulation for Massive MIMO Systems}
\author{Ahmed Raafat, Adrian Agustin and Josep Vidal
 \\ 
\small \begin{tabular}{c} Dept. of Signal Theory and Communications,
Universitat Politecnica de Catalunya (UPC),
Barcelona, Spain.
 \\
\small Email: $\lbrace$ahmed.raafat, adrian.agustin, josep.vidal$\rbrace$@upc.edu \\
\end{tabular}}


\maketitle

\newpage
\begin{abstract}
In this paper, we consider the downlink of a  massive multiple-input-multiple-output (MIMO) single user transmission system operating in the millimeter wave outdoor narrowband channel environment.
We propose a novel receive spatial modulation architecture aimed to reduce the power consumption at the user terminal, while attaining a significant spectral efficiency and low bit error rate.
The energy consumption reduction is obtained through the use of analog devices (amplitude detector), which reduces the number of radio frequency chains and analog-to-digital-converters (ADCs).
The base station transmits spatial and modulation symbols per channel use. 
We show that the optimal spatial symbol detector is a threshold detector that can be implemented by using one bit ADC. 
We derive closed form expressions for the detection threshold at different signal-to-noise-ratio (SNR) regions.
We derive expressions for the average bit error probability in the presence and absence of the threshold estimation error showing that a small number of pilot symbols is needed.
A performance comparison  is done between the proposed system and fully digital MIMO
showing that a suitable constellation selection can reduce the performance gap. 
\end{abstract}
\section{Introduction}
The utilization of millimeter wave (mmWave) frequencies for cellular communication systems can allow high data rates 
links because of the availability of unused large bandwidths \cite{Conv}. 
Millimeter wave signals suffer from severe path loss; however, the small wavelengths permit using massive arrays at the base stations (BSs) and user terminals (UTs). 
Massive arrays compensate the path loss through their own beamforming gain and enhance the 
spectral efficiency by boosting the multiplexing gain. 
The UT circuit structure per antenna of the fully digital (FD) multiple-input-multiple-output (MIMO) system comprises radio-frequency (RF) chain and analog to digital converter (ADC). 
Unfortunately, these devices are expensive and power consuming especially at mmWave frequencies. 
Thus, massive mmWave MIMO systems based FD architecture suffer from cost and power consumption problems \cite{Conv}.

\let\thefootnote\relax\footnotetext{
The research leading to these results has been partially funded by the 5Gwireless project within the framework of H2020 Marie Sk\l{}odowska-Curie innovative
training networks (ITNs) and the project 5G\&B RUNNER-UPC 
(TEC2016-77148-C2-1-R) funded by AEI/FEDER-UE.
}

Several low complexity MIMO transceiver designs have been developed.
Hybrid analog and digital precoding techniques have been proposed 
with the aim of simplifying the architecture and reducing power consumption
\cite{hybrid1}
. The spectral efficiency of the hybrid systems approach the FD MIMO \cite{hybrid1}. 
Nevertheless, hybrid structures comprise large numbers of phase shifters (PSs), power splitters and power combiners for analog signal processing. These analog devices consume a large amount of power principally with large arrays 
\cite{Pow}. 

MIMO transceivers-based low precision ADCs have been studied to reduce the cost and power consumption \cite{L2}. 
Nonetheless, the use of low bit ADCs hamper the availability of perfect channel knowledge 
and entails a reduction of the spectral efficiency.

Spatial modulation (SM) schemes have been reported to reduce the number of RF chains and 
to achieve high throughput. 
In SM, the transmit antennas are exploited in transmitting extra information where part of the input data bits are used to select the set of active transmit antennas and
the UT detects this set \cite{GSM1}. However, SM techniques suffer from small antennas gain because most of the transmit antennas are silent. Further, the UT cannot detect the set of active transmit antennas precisely if the 
channel vectors are correlated. 

In contrast, the receive SM (RSM) techniques exploit the receive antennas to transmit more information \cite{rsm}.
However, conventional RSM methods use FD receiver architecture and can suffer from high performance degradation under 
low rank channel matrices. Thus, traditional SM/RSM methods are not convenient for mmWave systems. 
In \cite{Marco} a low complexity RSM system is introduced for indoor line of sight mmWave propagation.
Unlike the previous RSM methods, this paper focuses on designing a new and simple RSM receiver
based novel spatial detection method aimed to
exploit the spatial dimension to transmit modulation symbols reliably.
Moreover, 
we show that the proposed RSM system works efficiently in outdoor narrowband mmWave channels.

We consider a massive MIMO single user operating in the mmWave outdoor narrowband channel environment. 
We present a RSM system where the BS transmits modulation symbol from $M$ size constellation and spatial symbol per channel use.
We show that the optimal spatial symbol detector is a threshold detector that can be implemented by using one bit ADC. 
We derive closed form expressions for the detection threshold at different signal-to-noise-ratio (SNR) regions showing that 
a simple threshold can be obtained at high SNR and its performance approaches the exact threshold. 
We derive expressions for the average bit error probability (ABEP) in the presence and absence of the threshold estimation error, and show that a small number of downlink pilot symbols is needed. 
Simulation results show that the performance of the proposed system with the appropriate constellation design approaches the 
FD MIMO system. 
\section{System Model}
\begin{figure}
\centering

\includegraphics[width=\linewidth,height=65mm]{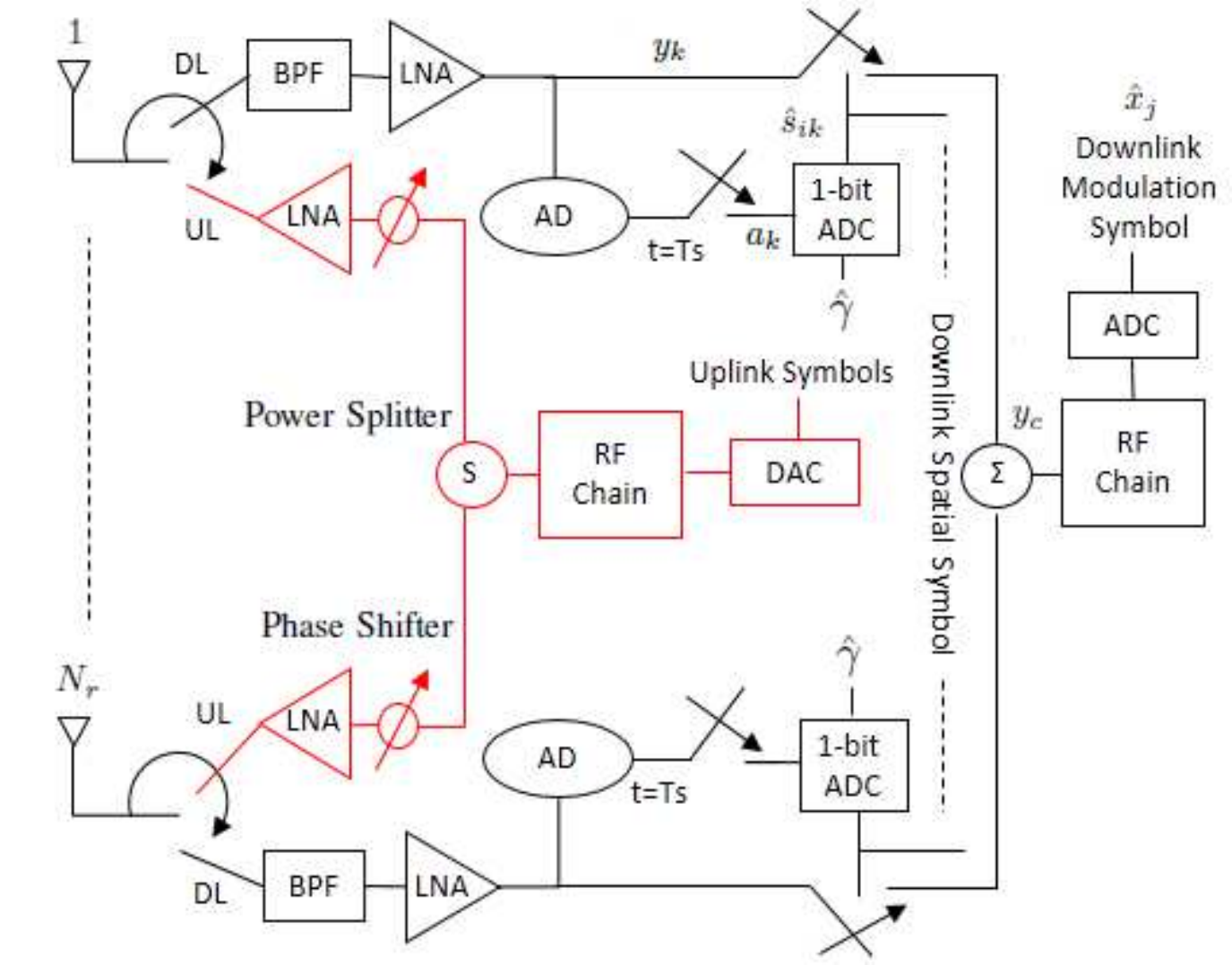}
\caption{Downlink (black) and uplink (red) MIMO UT circuits.}
\label{fig01}
\end{figure}
\subsection{Signal Model and Precoder Design}
We consider the downlink of a large MIMO single user that operates in 
the mmWave outdoor narrowband channel environment. 
The BS and UT are equipped with $N_{t}$ and $N_{r}$ antennas respectively. 
Based on the properties of mmWave propagation, we consider reciprocal propagation environment.
We exploit the channel reciprocity by considering time division duplex (TDD) system where the  
channel state information (CSI) is needed only at the BS. 
In Fig. \ref{fig01}, displayed in red, we consider a low complexity uplink UT circuit based on analog PSs. During the uplink training, the
UT sends pilot symbols so that the BS can 
acquire the CSI. 
This can be achieved, since the optimal training pilot symbols matrix for the least squares channel estimation 
can be selected as a discrete-fourier-transform (DFT) basis \cite{CE} that can be implemented by PSs.  
Since massive mmWave MIMO systems suffer from high path loss and antenna correlation,
antenna selection at the receiver is necessary. The BS selects the most suitable $N_{a}$ antennas based on the channel knowledge (assumed perfect in this paper).
After that, the BS informs the UT by the $N_{a}$ active receive antennas (ARA).
The BS transmits data vector to the ARA that comprises spatial and modulation symbols. The transmitted data vector 
can be written as 
\begin{equation}
\mathbf{x}_{i}^{j}=\sqrt{\alpha P}\mathbf{B}\mathbf{s}_{i}x_{j}
\label{eq1}
\end{equation}
where the spatial symbol $\mathbf{s}_{i} \in \mathbb{R}^{N_{a}\times 1}$ contains $N_{a}$ bits from the input data bits,
 $i \in \lbrace 1,...,2^{N_{a}}-1 \rbrace$, we assume that the all-zeros spatial symbol is not allowed, 
 the modulation symbol $x_{j}$ is a symbol from certain
 constellation with size $M$, $j \in \lbrace 1,...,M \rbrace$, 
 the number of transmitted bits per data vector are $(N_{a}+\log_{2}{M})$, 
 $\mathbf{B} \in \mathbb{C}^{N_{t}\times N_{a}}$ is the precoding matrix, $P$ is 
 the average transmit power and 
we adjust the transmitted power by a normalization factor $\alpha\approx\left(\Tr\left\{\!\mathbf{B}^{H}\mathbf{B}\!\right\}\right)^{-1} $ where $\text{Tr}\lbrace .\rbrace$ is the trace operator. The received signal vector 
is given by 
\begin{equation}
\mathbf{y}=\sqrt{\alpha P}\mathbf{H}\mathbf{B}\mathbf{s}_{i}x_{j}+\mathbf{n}
\label{eq3}
\end{equation}
where $\mathbf{H} \in \mathbb{C}^{N_{r}\times N_{t}}$ is the channel matrix and 
$\mathbf{n} \in \mathbb{C}^{N_{r}\times 1}$ is the generated noise vector where its coefficients are
independent and identically distributed (i.i.d) zero mean circularly symmetric complex Gaussian random variables and each has variance $\sigma^2$. 
Let us define the matrix $\mathbf{H}_{a}$ as the channel matrix from the BS to the selected ARAs. 
In order to direct the data vector to the ARA, we design the precoding matrix $\mathbf{B}$ as a zero forcing precoder 
that can be expressed as 
\begin{equation}
\mathbf{B}=\mathbf{H}_{a}^{H}\left(\mathbf{H}_{a}\mathbf{H}_{a}^{H}\right)^{-1}
\label{eq4}
\end{equation}

The received signal by the $k^{\text{th}}$ active antenna can be expressed as 
\begin{equation}
y_{k}= \sqrt{\alpha P}s_{ik}x_{j}+n_{k} 
\label{eq5}    
\end{equation}
where $s_{ik} \in\lbrace 0,1\rbrace$. For a given $N_{a}$,
the BS selects the best $N_{a}$ ARA such that the received power is maximized
that is, $\alpha$ is maximized, where $\alpha^{-1}=\Tr\left\{\!(\mathbf{H}_{a}\mathbf{H}_{a}^{H})^{-1}\!\right\}$. The maximization of $\alpha$ can be done through the exhaustive search.
A simple method for solving a similar problem was mentioned in \cite{cor}.
Low complexity algorithms that exploit the sparse nature of the outdoor mmWave channel to maximize $\alpha$ are left for future work. We assume that the UT can be successfully informed by the ARA through a control channel. Imperfect 
detection to the ARA at the UT is a topic for future research.

According to equation (\ref{eq5}), 
all the ARA associated to a spatial bit $s_{ik}=1$ receive the same modulation symbol.
The $k^{\text{th}}$ ARA is connected to RF switch that passes the signal only if the estimated spatial bit $\hat{s}_{ik}=1$. 
All of the signals that pass through the switches are combined. After that, the combined signal passes through 
RF chain and a single ADC to detect the modulation symbol. The combined signal is given by
\begin{equation}
y_{c}=\sum_{k=1}^{N_{a}}s_{ik}\hat{s}_{ik}\sqrt{\alpha P}x_{j}+\hat{s}_{ik}n_{k}
\label{eq101}
\end{equation}
where the optimal $\hat{s}_{ik}$ detection scheme is shown in section \ref{a}. In order to avoid the case $y_{c}=0$, 
we assume that the all-zeros transmit spatial symbol is not allowed. 

The downlink circuit in Fig. \ref{fig01} (displayed in black lines) shows that each receive antenna is connected to RF amplitude detector (AD). The AD 
works efficiently with high sensitivity and negligible power consumption at mmWave frequencies \cite{AD}. The signal provided by the AD at the $k^{\text{th}}$ active antenna is
\begin{equation}
\medmuskip=2mu   
\thickmuskip=3mu 
a_{k}=\sqrt{\left( \sqrt{\alpha P}s_{ik}x_{jI}+n_{kI}\right)^{2}+\left( \sqrt{\alpha P}s_{ik}x_{jQ}+n_{kQ}\right)^{2}}
\label{eq6}    
\end{equation}
where the indices $I$ and $Q$ represent the in-phase and quadrature components. 
The probability density function (PDF) of the received amplitude at the $k^{\text{th}}$ active antenna follows either a Rice or 
Rayleigh distributions \cite{book}. The PDF of the received amplitude can be expressed as 
\begin{equation}
f(a_{k})= \begin{cases}
\frac{2a_{k}}{\sigma^{2}}e^{-\frac{a_{k}^{2}+\alpha P}{\sigma^{2}}}\text{I}_{0}\left(
\frac{2a_{k}\sqrt{\alpha P}}{\sigma^{2}}\right)  &\text{if}\quad s_{ik}=1\\
\frac{2a_{k}}{\sigma^{2}}e^{-\frac{a_{k}^{2}}{\sigma^{2}}}&\text{if}\quad s_{ik}=0 
\end{cases}
\label{eq12}    
\end{equation}
where $\text{I}_{0}(x)$ is the zero order modified Bessel function of the first kind.
\subsection{Channel Model}
Millimeter wave channels have limited scattering clusters due to the high path loss. Moreover, using large arrays increases the antennas correlation. Therefore, we choose for system evaluation the narrowband clustered channel model that is widely used for outdoor mmWave channels \cite{hybrid1}. 
In this model, the channel matrix can be expressed as 
\begin{equation}
\medmuskip=2mu   
\thickmuskip=3mu 
\mathbf{H}=\sqrt{\frac{N_{t}N_{r}}{N_{c}N_{y}}}\sum\limits_{i=1}^{N_{c}}\sum\limits_{l=1}^{N_{y}}
g_{il}\Lambda_{r}(\theta_{il}^{r})\Lambda_{t}(\phi_{il}^{t})\mathbf{v}_{r}(\theta_{il}^{r})\mathbf{v}_{t}(\phi_{il}^{t})^{H}
\label{eq7}
\end{equation}
where $N_{c}$ represents number of scattering clusters, $N_{y}$ is the number of rays per cluster, 
$g_{il}$ is the complex gain, 
$\theta_{il}^{r}, \phi_{il}^{t}$  are the elevation and azimuth angles of arrivals and departures, $\Lambda_{r}(\theta_{il}^{r}), \Lambda_{t}(\phi_{il}^{t})$ are the receive and transmit directional antennas gains, $\mathbf{v}_{r}(\theta_{il}^{r}), \mathbf{v}_{t}(\phi_{il}^{t})$ are the receive and transmit array response vectors.
 
We consider uniform linear arrays where the $N$ antennas normalized response vector can be expressed as 
\begin{equation}
\mathbf{v}(\phi)=\frac{1}{\sqrt{N}}\left[1,e^{jkd\sin(\phi)},...,e^{j(N-1)kd\sin(\phi)}\right]^{T}
\label{eq8}
\end{equation} 
where $k=\frac{2\pi}{\lambda}$ and $d$ is the inter-antenna spacing. The directional antennas gain can be 
expressed as 
\begin{equation}
\Lambda(\phi)= \begin{cases}
1 & \forall \phi \in [\phi_{min},\phi_{max}]\\
0 & else
\end{cases}
\label{eq9}
\end{equation}
where $\phi_{min}$ and $\phi_{max}$ determine the transmission sector angle.
\subsection{Power Consumption}
The power consumption of the significant UT circuit components can  
be expressed in terms of reference power \cite{Pow} as

\noindent\begin{minipage}{0.33\linewidth}
\begin{equation} \nonumber
  P_{\text{RF chain}}=2P_{\text{ref}} \label{eq220}
\end{equation}
\end{minipage}%
\begin{minipage}{0.33\linewidth}
\begin{equation}\nonumber
   P_{\text{ADC}}=10P_{\text{ref}} \label{eq230}
\end{equation}
\end{minipage}%
\begin{minipage}{0.33\linewidth}
\begin{equation}\nonumber
 P_{\text{SW}}=0.25P_{\text{ref}} \label{eq2300}
\end{equation}
\end{minipage}\par\vspace{\belowdisplayskip}
\noindent\begin{minipage}{0.33\linewidth}
\begin{equation} \nonumber
 P_{\text{LNA}}=P_{\text{ref}} \label{eq2200}
\end{equation}
\end{minipage}%
\begin{minipage}{0.34\linewidth}
\begin{equation}\nonumber
 P_{\text{BB}}=N_{\text{RF}}P_{\text{ref}} \label{eq2400}
\end{equation}
\end{minipage}%
\begin{minipage}{0.33\linewidth}
\begin{equation}
 P_{\text{PS}}=1.5P_{\text{ref}} \label{eq230x}
\end{equation}
\end{minipage}\par\vspace{\belowdisplayskip}
\vspace*{1mm}
\hspace{-3.5mm}where $N_{\text{RF}}$ is the number of the RF chains, $P_{\text{SW}}$, $P_{\text{LNA}}$ and $P_{\text{BB}}$ are the switch, low noise amplifier and the base band power consumption respectively.
The uplink and downlink power consumption at the UT with the proposed system $P_{\text{P}}$ and with FD MIMO $P_{\text{FD}}$ can be expressed as 
\begin{eqnarray}
P_{\text{P}}=&\!\!\!\!\!\!\!\!\!\!\!\!\!\!\!\!\!\!\!\!\!\!\!\!\!\!\!\! N_{r}\!\!\!\!\!\!\!\!\!\!\!\!\!\!\!\!\!\!\!\!&\!\!\!\!\!\!\!\!\left(2P_{\text{LNA}}+P_{\text{PS}}+P_{\text{SW}}\right)+2\left(P_{\text{RF chain}}+P_{\text{ADC}}\right)+P_{\text{BB}}\nonumber \\
&P_{\text{FD}}\!\!\!\!\!&=2N_{r}\left(P_{\text{LNA}}+P_{\text{RF chain}}+P_{\text{ADC}}\right)+P_{\text{BB}}
\label{eqd1}
\end{eqnarray}

The power consumption ratio $P_{\text{P}}/P_{\text{FD}}\approx (0.14+\frac{0.9}{N_{r}})$. At
60 GHz and 500 MHz bandwidth systems, $P_{\text{ref}}=20$mW \cite{Pow}, $N_{r}=16$, $P_{\text{P}}=1700$mW and $P_{\text{FD}}=8640$mW, the hybrid MIMO architecture consumes $8000$mW \cite{Pow}.  
\section{Symbol Detection}\label{a}
\subsection{Spatial Symbol Detection}
In order to recover the spatial symbol, each ARA is connected to AD and one bit ADC 
as illustrated in Fig. \ref{fig01}. The AD measures the amplitude of the received signal and it is compared to 
a predefined threshold at the ADC. 
Thus, the output signals from the ADCs represent the spatial symbol. In the spatial symbol detection, we will show that both per antenna detection and joint detection lead to the same results.
\vspace{-6mm}
\subsection{Separate Spatial Symbol Detection}
We consider maximum likelihood (ML) detector per ARA to decide if the received spatial bit is one or zero.
The detection problem per the $k^{th}$ antenna can be formulated as 
\begin{eqnarray}
&f(a_{k}\vert s_{ik}=1)\underset{\hat{s}_{ik}=0}{\overset{\hat{s}_{ik}=1}{\gtrless}}f(a_{k}\vert s_{ik}=0) \\ 
&\frac{2a_{k}}{\sigma^{2}}e^{-\frac{a_{k}^{2}+\alpha P}{\sigma^{2}}}\text{I}_{0}\left(
\frac{2a_{k}\sqrt{\alpha P}}{\sigma^{2}}\right) \underset{\hat{s}_{ik}=0}{\overset{\hat{s}_{ik}=1}{\gtrless}} \frac{2a_{k}}{\sigma^{2}}e^{-\frac{a_{k}^{2}}{\sigma^{2}}}\\ 
&e^{-\frac{\alpha P}{\sigma^{2}}}\text{I}_{0}\left(
\frac{2a_{k}\sqrt{\alpha P}}{\sigma^{2}}\right) \underset{\hat{s}_{ik}=0}{\overset{\hat{s}_{ik}=1}{\gtrless}} 1
\label{eq13}    
\end{eqnarray}

According to the problem in (\ref{eq13}), the estimated spatial bit for the $k^{\text{th}}$ antenna can be expressed as 
\begin{equation}
\hat{s}_{ik}= \begin{cases}
1 & \text{if} \ a_{k}>\gamma\\
0& \text{if} \ a_{k}<\gamma
\end{cases}
\label{eq14}    
\end{equation}
where $\gamma$ is a threshold that results from solution of the problem in (\ref{eq13}). In the following, we present three ways to determine $\gamma$
based on the received SNR. 
\subsubsection{Exact Threshold}
We can obtain the exact value of the threshold directly by solving (\ref{eq13}) numerically
at the expenses of increasing
the UT circuit complexity. 
\subsubsection{Moderate SNR Approximation (MSA)}
At moderate SNR, we can approximate $\text{I}_{0}(x)\approx \frac{e^{x}}{\sqrt{2\pi x}}$ in problem (\ref{eq13}) and calculate the threshold by solving the following equation  
\begin{equation}
\frac{\sigma^{2}}{4\pi\gamma \sqrt{\alpha P}}e^{\frac{4\gamma\sqrt{\alpha P}-2\alpha P}{\sigma^{2}}}=1
\label{eq15}
\end{equation}
putting it in the form $xe^{x}=c$ as 
\begin{equation}
\frac{-4\gamma \sqrt{\alpha P}}{\sigma^{2}}e^{\frac{-4\gamma \sqrt{\alpha P}}{\sigma^{2}}}=-\frac{1}{\pi}e^{-\frac{2\alpha P}{\sigma^{2}}}
 \label{eq16}
\end{equation}

The solution of the equation in (\ref{eq16}) can be given by 
\begin{equation}
\gamma = \frac{-\sigma^{2}}{4\sqrt{\alpha P}} \text{W}_{-1}\left(-\frac{1}{\pi}e^{-2\frac{\alpha P}{\sigma^{2}}} \right)
\label{eq17}
\end{equation}
where $\text{W}_{-1}(x)$ is one of the main branches of the Lambert W function \cite{LW}. Nevertheless, we have to calculate the threshold in (\ref{eq17})
numerically and this leads to increase in the UT circuit complexity. 
\subsubsection{High SNR Approximation (HSA)}
At high SNR, the left hand side of equation (\ref{eq15}) takes either infinity or zero values based on the 
sign of the term in the exponential power. Therefore, we can obtain a simple threshold that can be expressed as 
\begin{equation}
\gamma=\frac{1}{2}\sqrt{\alpha P}
\label{eq18}
\end{equation} 

Since the received modulation symbol may come from non-constant amplitude constellation, the exact, MSA and HSA thresholds
should be designed based on the minimum received constellation symbol amplitude. This can be achieved by replacing the 
average power $P$ by the minimum power $P_{min}=\beta P$ in thresholds expressions where $\beta$ depends on the constellation.
\vspace{-2mm}     
\subsection{Joint Spatial Symbol Detection}
In order to jointly detect the spatial symbol bits, we apply the ML detector based on the received 
amplitudes from all of the active antennas. 
Since the received amplitudes are i.i.d, their joint PDF can be expressed as 
\begin{equation}
f(\mathbf{a})=\prod_{k=1}^{N_{a}}f(a_{k})
\label{eq19}
\end{equation}

The joint ML detection problem can be formulated as 
\begin{equation}
\hat{\mathbf{s}}_{i}= \arg\max_{\mathbf{s}_{i}}\lbrace f(\mathbf{a}\vert \mathbf{s}_{i})\rbrace 
\label{eq20}    
\end{equation}

For the sake of simplicity, let us consider the case when the number of the ARA is two. In this case, 
the joint ML detection problem can be expressed as 
\begin{equation}
\hat{\mathbf{s}}_{i}= \arg\max\lbrace f_{01},f_{10},f_{11}\rbrace 
\label{eq21}    
\end{equation}
where $f_{nm}=f_{\mathbf{A}}\left(\mathbf{a}\vert s_{i1}=n, s_{i2}=m\right)$. 

As an illustrative example, we decide symbol $[0\hspace{1mm}1]^{T}$
if the following conditions are satisfied
\begin{eqnarray}
  f_{01}&\!\!\!\!=\!\!\!\!&f_{00}f_{2}>f_{10}=f_{00}f_{1} \label{eq22}\\
  f_{01}&\!\!\!\!=\!\!\!\!&f_{00}f_{2}>f_{11}=f_{00}f_{1}f_{2} \label{eq23}
\end{eqnarray}
where $f_{k}$ can be given by 
\begin{equation}
  f_{k}=e^{-\frac{\alpha P}{\sigma^{2}}}\text{I}_{0}\left(
\frac{2a_{k}}{\sigma^{2}}\sqrt{\alpha P}\right) \label{eq25}
\end{equation}

Inequalities in (\ref{eq22}) and (\ref{eq23}) 
imply that $a_{1}<\gamma$ and $a_ {2}>\gamma$ respectively. See equations (\ref{eq13}) and (\ref{eq14}) for more illustration. The analysis of the joint detection can be extended for any number of ARA. Therefore, 
the joint detection results are equal to the per antenna detection.
\section{Downlink Training}
The BS has to send pilot symbols to allow the UT estimates the detection threshold.
We consider that the entries of the transmitted spatial symbol are all ones.

In order to estimate the detection threshold, the UT estimates the average received signal amplitude and the noise level from the outputs of the ADs connected to the ARA. The joint PDF of 
the received amplitudes can be expressed as 
\begin{equation}
f(\mathbf{a}\vert \mathbf{1}_{a})=\prod_{k=1}^{N}
\frac{2a_{k1}}{\sigma^{2}}e^{-\frac{a_{k1}^{2}+\vartheta^{2}}{\sigma^{2}}}\text{I}_{0}\left(
\frac{2a_{k1}\vartheta}{\sigma^{2}}\right)
\label{eq26}
\end{equation}
where $\mathbf{1}_{a}$ is all-ones spatial symbol, $\vartheta=\sqrt{\alpha P}$, $a_{k1}$ is the measured amplitude at the $k^{\text{th}}$ active 
antenna according to $s_{ik}=1$, $N=N_{p}N_{a}$ and $N_{p}$ is the number of pilot symbols. We design the amplitude estimator $\hat{\vartheta}$ so as to maximize 
\begin{equation}
\medmuskip=2mu   
\thickmuskip=3mu 
\log{f(\mathbf{a}\vert \mathbf{1}_{a})}=
\sum_{k=1}^{N}\log{\left(\frac{2a_{k1}}{\sigma^{2}}\right)}-\frac{a_{k1}^{2}+\vartheta^{2}}{\sigma^{2}}
+\log{\text{I}_{0}\left(
\frac{2a_{k1}\vartheta}{\sigma^{2}}\right)}
\label{eq27}
\end{equation}  

By using the fact that $\text{I}_{0}(x)=\frac{e^{x}}{\sqrt{2\pi x}}$ for large $x$, we can simplify equation (\ref{eq27}) as 
\begin{equation}
\log{\text{I}_{0}\left(
\frac{2a_{k1}\vartheta}{\sigma^{2}}\right)}=\frac{2a_{k1}\vartheta}{\sigma^{2}}-\frac{1}{2}
\log{\frac{4\pi a_{k1}\vartheta}{\sigma^{2}}} 
\label{eq28}
\end{equation}

In order to find the ML amplitude estimator $\hat{\vartheta}_{ML}$, we solve 
the problem, $\frac{\partial}{\partial \vartheta}\log{f(\mathbf{a}\vert \mathbf{1}_{a})}=0$, that can be expressed as
\begin{equation}
\sum_{k=1}^{N}\left(\frac{-2\vartheta}{\sigma^{2}}+\frac{2a_{k1}}{\sigma^{2}}-\frac{1}{2\vartheta}
\right)
=0
\label{eq30}
\end{equation} 

From (\ref{eq30}), the $\hat{\vartheta}_{ML}$ can be expressed as 
\begin{equation}
\hat{\vartheta}_{ML}=\frac{\sum_{k=1}^{N}a_{k1}}{2N}+\frac{1}{2}\sqrt{\left(\frac{\sum_{k=1}^{N}a_{k1}}{N}\right)^{2}-\sigma^{2}}
\label{eq32}
\end{equation}


The ML estimator of the noise variance $\hat{\sigma}^{2}_{ML}$ can be obtained by solving 
$\frac{\partial}{\partial \sigma^{2}}\log{f(\mathbf{a}\vert \mathbf{1}_{a})}=0$ as
\begin{equation}
\hat{\sigma}^{2}_{ML}=\frac{2}{N}\sum_{k=1}^{N}\left(a_{k1}-\vartheta\right)^{2}
\label{eq33}
\end{equation}

By solving the equations in (\ref{eq32},\ref{eq33}) simultaneously, a closed form expression 
for the $\hat{\vartheta}_{ML}$ can be given as 
\begin{equation}
\hat{\vartheta}_{ML}=
\frac{2\sum_{k=1}^{N}a_{k1}}{3N}+\frac{1}{3}\sqrt{4\left(\frac{\sum_{k=1}^{N}a_{k1}}{N}\right)^{2}\!\!-
\frac{3}{N}\sum_{k=1}^{N}a_{k1}^{2}}
\label{eq34x}
\end{equation}

\section{Error Analysis}
We evaluate the proposed system performance based on the ABEP that 
can be expressed as 
\begin{equation}
\text{ABEP}=\frac{N_{a}P_{es}+\log_{2}{M}P_{em}}{N_{a}+\log_{2}{M}}
\label{eqv1}
\end{equation}
where $P_{es}$ and $P_{em}$ are the spatial and modulation bit error probabilities respectively that can be given by 
\begin{equation}
 P_{es}=0.5\left(P_{1}+P_{0}\right)
\label{eqv22}
\end{equation}
where $P_{1}=\text{Pr}\left(a_{k1}<\gamma \right)$ and $P_{0}= \text{Pr}\left(a_{k0}>\gamma \right)$. 
\begin{equation}
\medmuskip=1mu   
\thickmuskip=2mu 
P_{em}=\sum_{i=1}^{2^{N_{a}}-1}\sum_{n=1}^{2^{N_{a}}}\text{BEP}\left(x_{j}\in \mathcal{C}_{M}\vert \hat{\mathbf{s}}_{n}, \mathbf{s}_{i} \right)
\text{Pr}\left(\hat{\mathbf{s}}_{n}\vert \mathbf{s}_{i}\right)
\text{Pr}\left(\mathbf{s}_{i}\right)
\label{eqv23}
\end{equation}
where $\text{BEP}\hspace{0.5mm}(x_{j}\in \mathcal{C}_{M})$ is the bit error probability of $M$-quadrature-amplitude-modulation (M-QAM), 
$M$-phase-shift-keying (M-PSK) or $M$-amplitude-phase-shift-keying (M-APSK)
with rings ratio $r$ \cite{e1}. The probabilities in equation (\ref{eqv23}) can be expressed as
\begin{equation}
\text{BEP}\left(x_{j}\in \mathcal{C}_{M}\vert \hat{\mathbf{s}}_{n}, \mathbf{s}_{i} \right)
=\text{BEP}\left( x_{j} \in \mathcal{C}_{M}\vert \mathrm{SNR_{c}}\right)
\label{ah5}
\end{equation}
\begin{equation}
\mathrm{SNR_{c}}=\frac{(b_{in}^{11})^{2}}{b_{in}^{11}+b_{in}^{01}}\frac{\alpha P}{\sigma^{2}}\!\!\!\!\!\!\!\!\!
\label{ah7}
\end{equation}
\begin{equation}
\text{Pr}\left(\hat{\mathbf{s}}_{n}\vert \mathbf{s}_{i}\right)=P_{1}^{b_{in}^{10}}(1-P_{1})^{b_{in}^{11}}P_{0}^{b_{in}^{01}}(1-P_{0})^{b_{in}^{00}}
\label{ah6}
\end{equation}
\begin{equation}
\text{Pr}\left(\mathbf{s}_{i}\right)=\frac{1}{2^{N_{a}}-1}
\label{ah8}
\end{equation}
where $b_{in}^{(m_{1}m_{2})}$ is the number of entries in $\mathbf{s}_{i}$ that each one of them equals $m_{1}$ and its corresponding entry in $\hat{\mathbf{s}}_{n}$ equals $m_{2}$ and $\sum_{m_{1},m_{2}\in \lbrace 0,1\rbrace}b_{in}^{(m_{1}m_{2})}=N_{a}$. 
\subsection{Perfect Threshold}
If the UT knows the threshold $\gamma$ perfectly, the probabilities $P_{1}$ and $P_{0}$
can be expressed by the cumulative density function of Rice and Rayleigh distributions \cite{book} as
\begin{eqnarray} \nonumber
P_{1}&\!\!\!\!=\!\!\!\!&1-Q_{1}\left(\frac{1}{\sigma}\sqrt{2\alpha P},
\frac{1}{\sigma}\sqrt{2}\gamma \right) \\
 P_{0}&\!\!\!\!=\!\!\!\!&e^{-\frac{\gamma^{2}}{\sigma^{2}}} \label{eqv27}
\end{eqnarray}
where $Q_{1}(x)$ is the first order Marcum Q-function.
\subsection{Estimated HSA Threshold}
The ML estimator is asymptotically gaussian 
so we consider the 
estimated HSA threshold 
$\hat{\gamma}\sim \mathcal{N}(\mu_{\hat{\gamma}},\,\sigma^{2}_{\hat{\gamma}})$ where $\mu_{\hat{\gamma}}$ and $\sigma^{2}_{\hat{\gamma}}$
are the mean and variance of $\hat{\gamma}$.
Since the received amplitudes are positive,
we can express $P_{1}$ and $P_{0}$ as
\begin{eqnarray}\nonumber
 P_{1}&\!\!\!\!=\!\!\!\!&\text{Pr}\left(\frac{\hat{\gamma}}{a_{k1}}>1 \right)\\
 P_{0}&\!\!\!\!=\!\!\!\!&\text{Pr}\left(\frac{\hat{\gamma}}{a_{k0}}<1 \right) \label{eq38}
\end{eqnarray}

In order to find the probabilities in equation (\ref{eq38}), we put it in the following form
\begin{eqnarray}\nonumber
 P_{1}&\!\!\!\!=\!\!\!\!&\text{Pr}\left(t_{\delta,n,l}>\frac{\sigma}{\sigma_{\hat{\gamma}}} \right)\\
 P_{0}&\!\!\!\!=\!\!\!\!&\text{Pr}\left(t_{\delta,n}<\frac{\sigma}{\sigma_{\hat{\gamma}}} \right) \label{eq39}
\end{eqnarray}
where $t_{\delta,n}$ is Non-central t distribution, $t_{\delta,n,l}$ is Doubly-non-central t distribution \cite{dis}, 
$n$ is the degrees of freedom, $\delta, l$ are the non-centrality parameters,
$\delta=\frac{\mu_{\hat{\gamma}}}{\sigma_{\hat{\gamma}}}$, n=2 and $l=\frac{2\alpha P}{\sigma^{2}}$. A closed form expressions 
for $P_{1}$ and $P_{0}$ with threshold estimation error can be given as 
\begin{eqnarray}\nonumber
P_{1}&\!\!\!\!=\!\!\!\!&1-T_{\delta,n,l}\left(\frac{\sigma}{\sigma_{\hat{\gamma}}}\right)\\
P_{0}&\!\!\!\!=\!\!\!\!&T_{\delta,n}\left(\frac{\sigma}{\sigma_{\hat{\gamma}}}\right) \label{eq40}
\end{eqnarray}
where $T_{\delta,n}$ and $T_{\delta,n,l}$ are the cumulative density functions of Non-central and 
Doubly-non-central t distributions \cite{dis}. 

In order to determine the probabilities in equation (\ref{eq40}), we need the mean and the variance of $\hat{\gamma}$. 
By relating equations (\ref{eq34x}) and (\ref{eq18}),
the estimated HSA threshold $\hat{\gamma}$ is 
 \begin{equation}
\hat{\gamma}=\frac{1}{2}\hat{\vartheta}_{ML}
\label{eq41}
\end{equation}

From the asymptotic properties of the ML estimator \cite{estimation}, the mean and variance of $\hat{\gamma}$  
can be expressed as 
\begin{equation}
\mu_{\hat{\gamma}}=\frac{1}{2}\vartheta
\label{eqahm1}
\end{equation}
\begin{equation}
\sigma^{2}_{\hat{\gamma}}=\frac{1}{4}\left[\mathbf{I}^{-1}_{\theta}\right]_{11}
\label{eqahm2x}
\end{equation}
where $\mathbf{I}_{\theta}$ is a $2\times 2$ fisher information matrix \cite{estimation} whose elements can be expressed as 
\begin{eqnarray} \nonumber
\left[\mathbf{I}_{\theta}\right]_{11}&\!\!\!=\!\!\!&-\Exp{\frac{\partial^2 f(\mathbf{a}\vert \mathbf{1}_{a})}{\partial \vartheta^2}}=\frac{2N}{\sigma^{2}}-\frac{N}{2\vartheta^{2}}\\ \nonumber
\left[\mathbf{I}_{\theta}\right]_{12}&\!\!\!=\!\!\!&\left[\mathbf{I}_{\theta}\right]_{21}\!\!=
-\Exp{\frac{\partial^2 f(\mathbf{a}\vert \mathbf{1}_{a})}{\partial \vartheta \partial \sigma^{2}}}
=\frac{2N}{\sigma^{4}}\left(\mu_{1}-\vartheta \right)\\ \nonumber
\left[\mathbf{I}_{\theta}\right]_{22}&\!\!\!=\!\!\!&
-\Exp{\frac{\partial^2 f(\mathbf{a}\vert \mathbf{1}_{a})}{\partial \sigma^2 \partial \sigma^2}}\\
&\!\!\!=\!\!\!&
\frac{2N}{\sigma^{6}}\left(\mu_{2}\!\!+\!\vartheta^{2}\!\!-\!\!2\vartheta\mu_{1}\right)\!-\!\!
\frac{N}{2\sigma^{4}}
\end{eqnarray}
where $\Exp{.}$ is the expectation operator, $\mu_{1}=\Exp{a_{k1}}$ and $\mu_{2}=\Exp{a_{k1}^{2}}$.
The inverse of the fisher information matrix can be expressed as 
\begin{equation}
\mathbf{I}_{\theta}^{-1}=\frac{1}{\left[\mathbf{I}_{\theta}\right]_{11}\left[\mathbf{I}_{\theta}\right]_{22}-\left[\mathbf{I}_{\theta}\right]_{12}^{2}}
  \begin{bmatrix}
    \left[\mathbf{I}_{\theta}\right]_{22} & -\left[\mathbf{I}_{\theta}\right]_{12}  \\
    -\left[\mathbf{I}_{\theta}\right]_{21} & \left[\mathbf{I}_{\theta}\right]_{11} 
  \end{bmatrix}
\label{eqah1x}
\end{equation} 

From equations (\ref{eqahm2x},\ref{eqah1x}), the variance $\sigma^{2}_{\hat{\gamma}}$ can be 
given as 
\begin{equation}
\sigma^{2}_{\hat{\gamma}}=
\frac{1}{4}
\frac{\left[\mathbf{I}_{\theta}\right]_{22}}{\left[\mathbf{I}_{\theta}\right]_{11}\left[\mathbf{I}_{\theta}\right]_{22}-\left[\mathbf{I}_{\theta}\right]_{12}^{2}}
\end{equation}
\vspace{-.5mm}
\section{Simulation Results}
In this section, we present simulation results that show the performance of the proposed system. In simulation environment, 
we consider that $g_{il}$ are i.i.d $\mathcal{CN}(0,\,\sigma^{2}_{g})$ where $\sigma_{g}^2$ is designed such that 
$\Exp{\Tr\left\{\!\mathbf{H}^{H}\mathbf{H}\!\right\}}=N_{t}N_{r}$, $N_{c}=8$, $N_{y}=10$ \cite{hybrid1},
the elevation and azimuth angles $(\theta_{il},\phi_{il})$ have Laplacian distributions with uniform 
random means $(\theta_{i},\phi_{i})$ and angular spreads $\sigma_{\theta}=\sigma_{\phi}=1$, $\text{SNR}=\frac{P}{\sigma^{2}}$, the width of the transmission angle is $50 \degree$ and the user has omni-directional antenna array. We compare the performance of the proposed system with singular value decomposition (SVD) based precoding and decoding of the FD MIMO system. 
We allocate the power at SVD in such a way that all of the activated modes achieve the same received SNR \cite{SV}. 

\begin{figure}[t!]
\centering
\includegraphics[width=\linewidth, height=60mm]{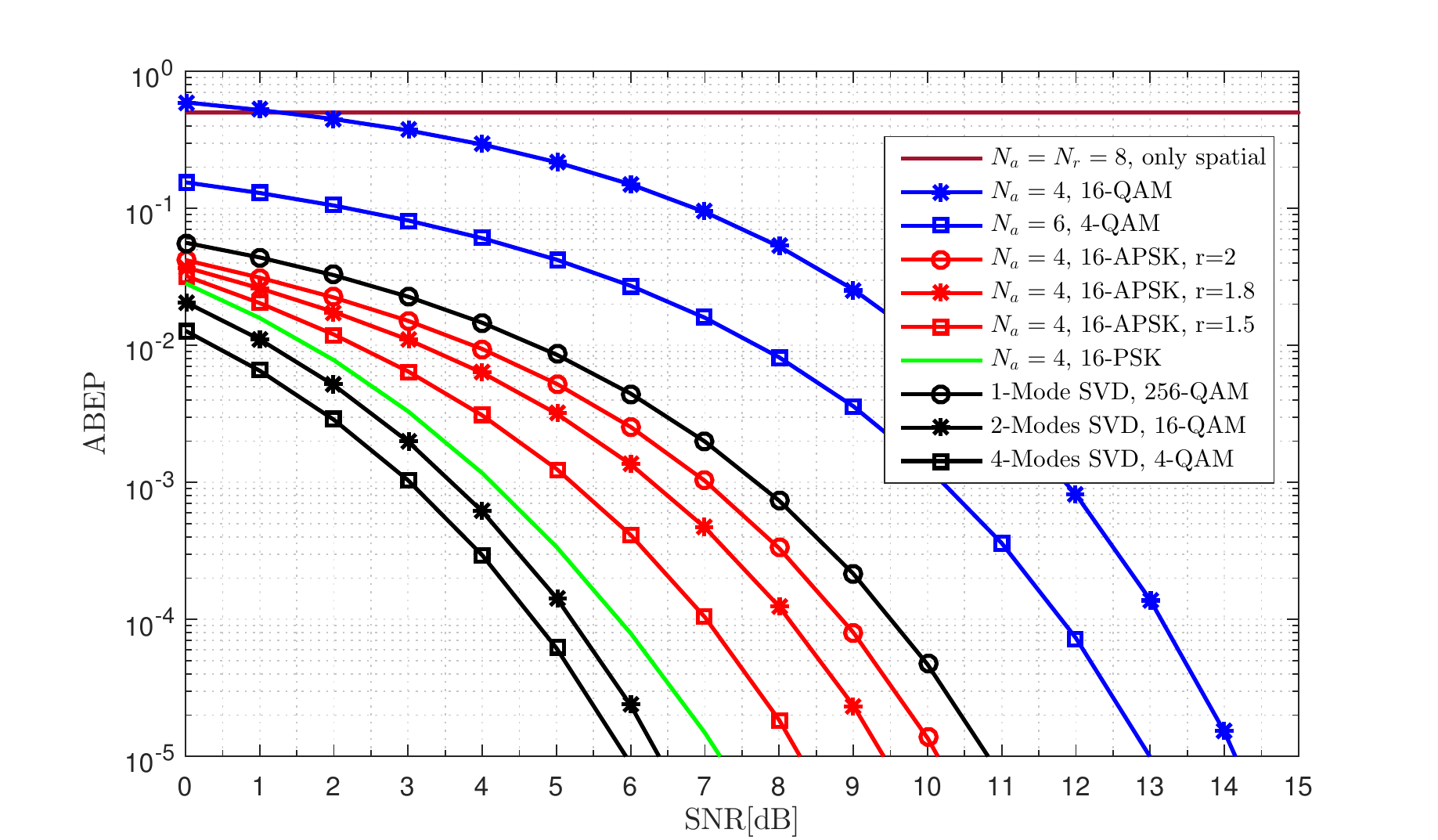}
\caption{Average bit error probability of the proposed system and FD MIMO
 at different constellation designs. 8 transmit bits are used in all cases.}
\label{fig22}
\end{figure}

Fig. \ref{fig22} shows the ABEP of the proposed system with exact threshold compared to FD MIMO at $32\times 8$ MIMO system. 
Clearly, the performance of the proposed system without receive antenna selection is 
inferior to that with antenna selection. Moreover, 
the proposed system performance with constant amplitude constellation and $4$ spatial bits approaches the FD MIMO.
The optimal values of spatial, modulation bits and constellation design to minimize the ABEP are topics for future research.

\begin{figure}[t!]
\centering
\includegraphics[width=\linewidth, height=60mm]{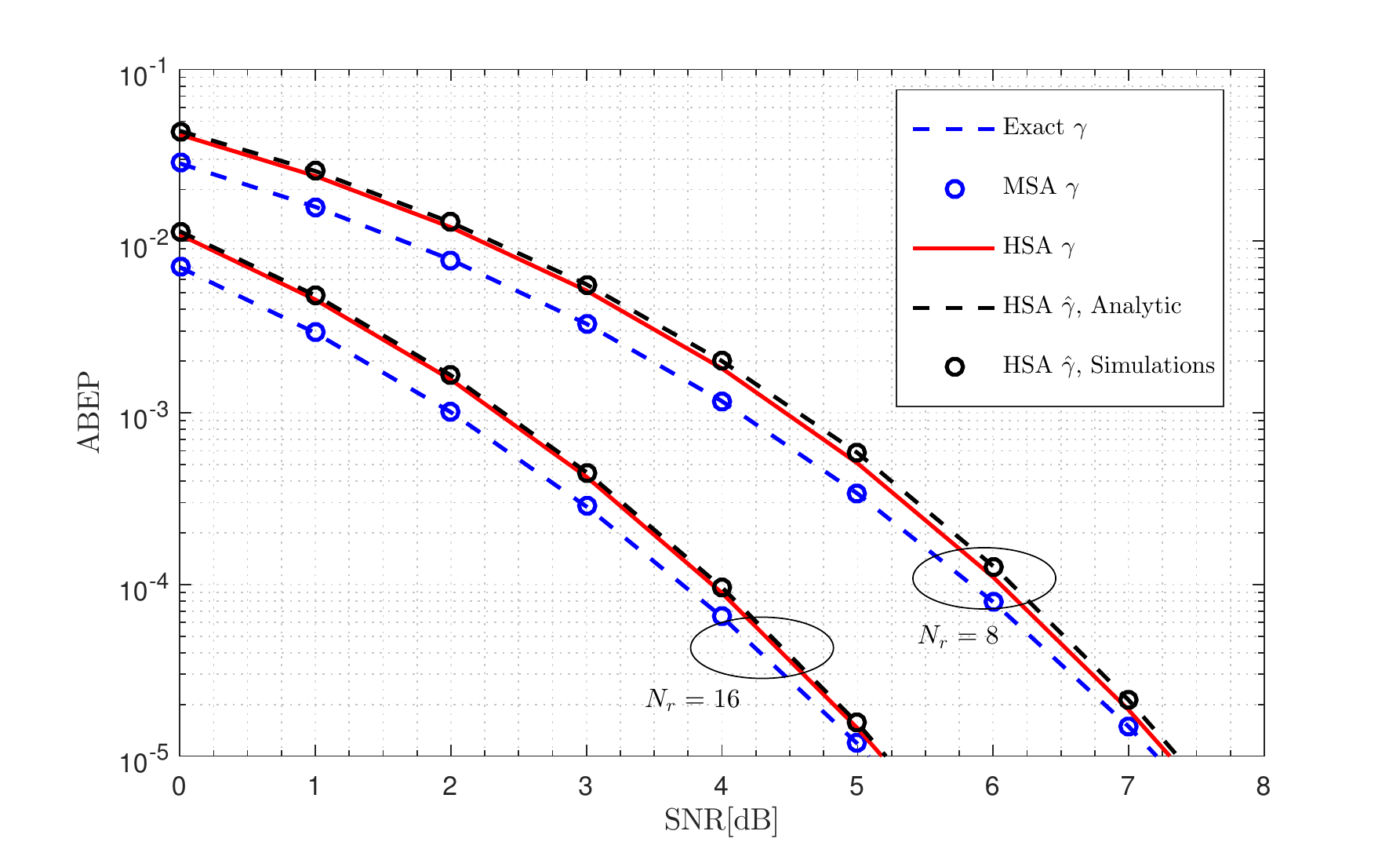}
\caption{Average bit error probability of the proposed system at $N_{t}=32$, $N_{a}=4$, 16-PSK, perfect and estimated thresholds.}
\label{fig33}
\end{figure}

Fig. \ref{fig33} represents the ABEP of the proposed system at different thresholds and different numbers of receive antennas. 
Applying HSA threshold leads to the lowest complexity and the performance gap is less than 1 dB with respect to the exact
threshold. The ABEP with threshold estimation error is very close to that with perfect threshold by using only one downlink pilot symbol. Increasing the number of receive antennas while $N_{a}$ is fixed boosts the receive antennas gain and as a result the ABEP is improved.   
\section{Conclusion}
In this paper, 
we have considered the downlink of a massive MIMO single user operating in the mmWave outdoor narrowband 
channel environment.
The proposed RSM architecture relies on one RF chain and one high resolution ADC at the UT 
that reduces the power consumption and can achieve high spectral efficiency as the ARA are exploited to transmit 
extra spatial symbols. 
The constellation choice affects the system performance as the detection threshold is designed based the 
minimum constellation symbol amplitude. Therefore, a constant amplitude constellation with balancing between number of spatial and modulation bits outperforms other constellation designs. The receive antenna selection is necessary for the proposed system to 
work efficiently. However, this paper considered exhaustive search algorithm to select the ARA.  
Low complexity algorithms for receive antenna selection
and extending the proposed system to allow higher order spatial symbols transmission through using multiple-bit ADCs at 
the UT are future work topics.
\bibliographystyle{IEEEbib}
\bibliography{IEEEabrv,refrences2}

\begin{thebibliography}{10}

\bibitem{Conv}
R.~W. Heath, N.~Gonzalez-Prelcic, S.~Rangan, W.~Roh, and A.~M. Sayeed,
\newblock ``An overview of signal processing techniques for millimeter wave
  {MIMO} systems,''
\newblock {\em IEEE Journal of Selected Topics in Signal Processing}, vol. 10,
  no. 3, pp. 436--453, Feb. 2016.

\bibitem{hybrid1}
O.~El~Ayach, S.~Rajagopal, S.~Abu-Surra, Z.~Pi, and R.~W. Heath,
\newblock ``Spatially sparse precoding in millimeter wave { MIMO} systems,''
\newblock {\em IEEE Transactions on Wireless Communications}, vol. 13, no. 3,
  pp. 1499--1513, March 2014.

\bibitem{Pow}
R.~M{\'e}ndez-Rial, C.~Rusu, N.~Gonz{\'a}lez-Prelcic, A.~Alkhateeb, and R.~W.
  Heath,
\newblock ``Hybrid {MIMO} architectures for millimeter wave communications:
  {Phase} shifters or switches?,''
\newblock {\em IEEE Access}, vol. 4, pp. 247--267, Jan. 2016.

\bibitem{L2}
J.~Mo and R.~W. Heath,
\newblock ``Capacity analysis of one-bit quantized {MIMO} systems with
  transmitter channel state information,''
\newblock {\em IEEE Transactions on Signal Processing}, vol. 63, no. 20, pp.
  5498--5512, Oct. 2015.

\bibitem{GSM1}
A.~Younis, N.~Serafimovski, R.~Mesleh, and H.~Haas,
\newblock ``Generalised spatial modulation,''
\newblock {\em 44rd Asilomar Conference on Signals, Systems and Computers}, pp.
  1498--1502, Nov. 2010.

\bibitem{rsm}
R.~Zhang, Lie-Liang Yang, and L.~Hanzo,
\newblock ``Generalised pre-coding aided spatial modulation,''
\newblock {\em IEEE Transactions on Wireless Communications}, vol. 12, no. 11,
  pp. 5434--5443, Nov. 2013.

\bibitem{Marco}
N.~S. Perovic, P.~Liu, M.~Di~Renzo, and A.~Springer,
\newblock ``Receive spatial modulation for los mmwave communications based on
  {TX} beamforming,''
\newblock {\em IEEE Communications Letters}, Dec. 2016.

\bibitem{CE}
M.~Biguesh and A.~B. Gershman,
\newblock ``Training-based {MIMO} channel estimation: a study of estimator
  tradeoffs and optimal training signals,''
\newblock {\em IEEE Transactions on Signal Processing}, vol. 54, no. 3, pp.
  884--893, March 2006.

\bibitem{cor}
L.~Dai, S.~Sfar, and K.~B. Letaief,
\newblock ``Receive antenna selection for {MIMO} systems in correlated
  channels,''
\newblock {\em IEEE International Conference on Communications}, vol. 5, pp.
  2944--2948, June 2004.

\bibitem{AD}
S.~Rami, W.~Tuni, and W.~R. Eisenstadt,
\newblock ``Millimeter wave {MOSFET} amplitude detector,''
\newblock {\em Topical Meeting on Silicon Monolithic Integrated Circuits in RF
  Systems (SiRF)}, pp. 84--87, Jan. 2010.

\bibitem{book}
M.~K. Simon,
\newblock {\em Probability distributions involving Gaussian random variables:
  {A} handbook for engineers and scientists},
\newblock Springer, May 2007.

\bibitem{LW}
D.~Barry, J-Y Parlange, L.~Li, H.~Prommer, C.~Cunningham, and F.~Stagnitti,
\newblock ``Analytical approximations for real values of the{ Lambert
  W-function},''
\newblock {\em Mathematics and Computers in Simulation}, vol. 53, no. 1, pp.
  95--103, Aug. 2000.

\bibitem{e1}
J.~W. Craig,
\newblock ``A new, simple and exact result for calculating the probability of
  error for two-dimensional signal constellations,''
\newblock {\em IEEE Military Communications Conference}, pp. 571--575, Nov.
  1991.

\bibitem{dis}
C.~Walck,
\newblock {\em Handbook on statistical distributions for experimentalists},
\newblock University of Stockholm Internal Report, 2007.

\bibitem{estimation}
H.~L. Van~Trees,
\newblock {\em Optimum {Array} {Processing}, { Detection}, {Estimation}, and
  {Modulation} {Theory}},
\newblock John Wiley \& Sons, 2004.

\bibitem{SV}
B.~Vrigneau, J.~Letessier, P.~Rostaing, L.~Collin, and G.~Burel,
\newblock ``Extension of the {MIMO} precoder based on the minimum {Euclidean}
  distance: a cross-form matrix,''
\newblock {\em IEEE Journal of selected topics in signal processing}, vol. 2,
  no. 2, pp. 135--146, April 2008.

\end{thebibliography}
\end{document}